\documentclass[apj,twocolumn,twocolappendix,numberedappendix,appendixfloats]{openjournal}
\usepackage{natbib}

\usepackage{graphicx}	
\usepackage{amsmath}	
\usepackage{amssymb}	
\usepackage[dvipsnames,svgnames,table]{xcolor}
\usepackage[normalem]{ulem} 
\usepackage{newtxtext,newtxmath}
\usepackage{booktabs}
\usepackage{gensymb}
\usepackage[
breaklinks=true,
hyperindex=true,
colorlinks=true,
linkcolor=Sepia,
citecolor=Purple]{hyperref}

\newcommand{\kms}{\mathrm{km}\ \mathrm{s}^{-1}}

\newcommand{\Msun}{\mathrm{M}_\odot}
\newcommand{\gpcc}{\mathrm{g}\ \mathrm{cm}^{-3}}

\definecolor{darkgreen}{rgb}{0.25,0.51,0.12}

\newcommand{\orcidauthor}[3]{\author{#2$^{#3}$\href{http://orcid.org/#1}{\includegraphics[height=1em]{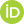}}}}

\begin{document}

\title[Gravity-driven mass inflow]{
What is the contribution of gravitational infall on the mass assembly of star-forming clouds? A case study in a numerical simulation of the interstellar medium.\vspace{-2em}}

\orcidauthor{0000-0003-3094-9674}{Noé Brucy}{*\ 1,2}
\orcidauthor{0000-0002-1424-3543}{Enrique Vázquez-Semadeni}{3}
\orcidauthor{0000-0002-2636-4377}{Tine Colman}{1}
\orcidauthor{0000-0002-4383-0202}{Jérémy Fensch}{1}
\orcidauthor{0000-0002-0560-3172}{Ralf S.\ Klessen}{2,4}

\thanks{$^*$ noe.brucy [at] ens-lyon.fr}

\affiliation{$^{1}$ ENS de Lyon, CRAL UMR5574, Universite Claude Bernard Lyon 1, CNRS, Lyon 69007, France}
\affiliation{$^{2}$ Universit\"{a}t Heidelberg, Zentrum f\"{u}r Astronomie, Institut f\"{u}r Theoretische Astrophysik, Albert-Ueberle-Str. 2, 69120 Heidelberg, Germany}
\affiliation{$^{3}$ Instituto de Radioastronom\'ia y Astrof\'isica, Universidad Nacional Aut\'onoma de M\'exico, Campus Morelia, Antigua Carretera a P\'atzcuaro \# 8701, Morelia, Michoac\'an, 58089, M\'exico}
\affiliation{$^{4}$ Universit\"{a}t Heidelberg, Interdisziplin\"{a}res Zentrum f\"{u}r Wissenschaftliches Rechnen, Im Neuenheimer Feld 225, 69120 Heidelberg, Germany
}

\begin{abstract}
Star formation in galaxies is a complex phenomenon occurring on a very wide range of scales, and molecular clouds are at the heart of this process. The formation of these structures and the subsequent collapse of the gas within them to form new stars remain unresolved scientific questions. In particular, the role and importance of gravity at between the disk scale height and prestellar cores (100 to 0.01 pc) are still topics of debate.
In this work, we conduct a case study examining the mass assembly and evolution of a giant molecular cloud complex in a numerical stratified-box simulation of the interstellar medium with photo-ionizing and supernova driving 
and resolving down to scales $\gtrsim 1$ pc and densities up to $10^3$ cm$^{-3}$.
By introducing tracer particles to precisely track the forces acting on the gas during its evolution towards and within the clouds, we are able to quantify how much of the mass inflow is driven by the self-gravity of the gas and the gravity from the stellar disk. We find that up to 20\% of the gas is gravity-driven at a scale of 100 pc, contributing 10\% of the inflow from the warm to the cold phase and 20\% from the cold phase to the individual molecular clouds, reaching up to 45\% inside the molecular gas, at densities $\gtrsim 400$ cm$^{-3}$. However, at the 100~pc scale, the contribution of gravity-driven gas on the linewidth is negligible. We conclude that the bulk of the gas motions assembling the clouds in our simulation are caused by the  supernova-driven supersonic turbulence, which 
results in locally convergent flows, with a small contribution from the stellar gravitational potential.
\end{abstract}


\maketitle



\section{Introduction}
\label{sec:intro}

Molecular clouds (MCs) are the densest regions in the Galactic interstellar medium (ISM) and the primary sites of star formation, even though they occupy only a small fraction of the ISM volume. The mechanisms responsible for assembling these clouds remain the subject of debate, particularly with regard to the relative roles of cloud-scale (or larger) gravitational potential \citep[e.g.,][]{vazquez-semadeniGlobalHierarchicalCollapse2019} versus large-scale coherent inertial motions driven by supernova (SN) explosions \citep[e.g.,][]{kimThreephaseInterstellarMedium2017,padoanOriginMassiveStars2020} or galactic-scale processes \citep{goldbaumMassTransportTurbulence2015,krumholzUnifiedModelGalactic2018,meidtModelOnsetSelfgravitation2020,brucyLargescaleTurbulentDriving2020}.

Turbulence-driven models of MCs and star formation assume that clouds are in near-virial equilibrium, where gravitational gradients are balanced by internal turbulent pressure gradients, acting in concert with other forces associated with radiation, cosmic rays, and or magnetic fields \citep[see, for example][]{mckeeFormationMassiveStars2003, VS+03, maclowControlStarFormation2004, krumholzFormationStarsGravitational2005, padoanStarFormationRate2011, hennebelleAnalyticalStarFormation2011, federrathStarFormationRate2012, klessenPhysicalProcessesInterstellar2016, 
hennebelleInefficientStarFormation2024}. This turbulence may be cascaded from large-scale motions such as SN shocks or generated internally by local stellar sources (jets, {\sc Hii} regions, winds, etc.). 
In this scenario, the supersonic linewidths of molecular tracers are interpreted primarily as turbulence.

\begin{table}
	\caption{Main notations used in the article.}
  \begin{center}
	\begin{tabular}{lp{4.5cm}}
	\toprule
	Notation &  Description \\
	\midrule
     $\vec{u} = u_x\vec{e}_x +  u_y\vec{e}_y +  u_z\vec{e}_z $ & vectors are noted with an arrow  \\
     $ u $ & If $\vec{u}$ is a vector, we write its norm $u$ \\
     \midrule
     $\rho$ & Density ($\gpcc$) \\
     $n$ & Number density of hydrogen atoms($\mathrm{cm}^{-3}$) \\
     $\vec{v}$ & Velocity \\
     $\vec{g}$ & Gravitational field (gas self-gravity and the vertical analytical field) \\
     $\vec{r}$ & Position vector in the frame of the minimum of the gravitational potential \\
     $m_i$ & mass of a gas tracer $i$\\
      \midrule
     $t_s= 74~\mathrm{Myr}$ &  Start of the studied time-span\\
     $t_e = 84~\mathrm{Myr}$  &  End of the studied time-span \\
     \midrule
     $\left<\vec{a}_{\mathrm{tot},i}\right>(t_s, t_e)$ & Averaged total acceleration of a tracer particle i between $t_s$ and $t_e$\\
     $\left<\vec{a}_{\mathrm{grav},i}\right>(t_s, t_e)$  & Averaged gravitational acceleration of a tracer particle i between $t_s$ and $t_e$ \\
     $\left<a_{\mathrm{other},i}\right>(t_s, t_e)$ & Averaged non-gravitational acceleration of a tracer particle i between $t_s$ and $t_e$ \\
    $\left<\vec{a}_{\mathrm{SG},i}\right>(t_s, t_e)$ & Averaged self-gravitational acceleration of a tracer particle i between $t_s$ and $t_e$ \\
    \midrule
    $\dot{\mathrm{M}}_\mathrm{H}(\rho, t_\mathrm{s}, t_\mathrm{e})$ & Averaged hierarchical mass flow of the tracers getting in denser environment than $\rho$ between $ t_\mathrm{s}$ and $ t_\mathrm{e}$\\
	\bottomrule
	\end{tabular} 
  \end{center}
	\label{tbl:notation}
\end{table}

In contrast, the Global Hierarchical Collapse (GHC) model \citep[see][]{vazquez-semadeniGlobalHierarchicalCollapse2019,vazquez-semadeniTurbulentSupportTS2024} proposes that even the largest-scale motions within and onto clouds are largely driven by gravity. This framework envisions a hierarchy of ``collapses within collapses,'' with gravity dominating at most levels, with turbulence still being responsible for producing nonlinear density fluctuations that act as seeds for guiding the collapse flow, and that can collapse faster than their parent structures because of their shorter free-fall times. 
Numerical simulations show that structures at all scales accrete from their parent structure \citep[e.g.,][]{VS+09,VS+10,ibanez-mejiaFeedingFallingGrowth2017, GS_VS20,kuffmeierRejuvenatingInfallCrucial2023,lebreuillySyntheticPopulationsProtoplanetary2024,ahmadFormationLowmassProtostars2024,mayerProtostellarDisksTheir2025},  due to the fact that denser structures are assembled by convergent flows gathering material from the more tenuous and diffuse phases of the ISM \citep[e.g.,][]{klessenAccretiondrivenTurbulenceUniversal2010, Hopkins2013, krumholzTurbulenceInterstellarMedium2016,padoanOriginMassiveStars2020,Forbes2023}. 
In the GHC model, turbulence contributes only $\sim$30--40\% of molecular linewidths, with the remainder due to gravitational contraction \citep{GG_VS20}.  
Although the GHC and turbulence-driven (TD) scenarios emphasize different dominant processes, they are not mutually exclusive: both involve initial turbulence seeding and subsequent gravitational collapse, but differ in the scales where gravity dominates.

Recent studies have sought to quantify mass inflows onto dense structures and determine the driving forces. For example, \cite{ibanez-mejiaGravityMagneticFields2022} performed zoom-in simulations of clouds from stratified-box models and found that gravitational acceleration is subdominant in diffuse gas but dominates in regions with number densities between $10^2$ and $10^3$ cm$^{-3}$, corresponding to mass densities of roughly $10^{-21}\ \gpcc$  \citep[see also][]{ibanez-mejiaGravitationalContractionSupernova2016, ibanez-mejiaFeedingFallingGrowth2017, gangulySILCCZoomDynamicBalance2024}. Similarly, \cite{appelWhatSetsStar2023} analyzed compression rates in dense clouds and concluded that gravity primarily drives motions at high densities. Observationally, measuring mass inflow is challenging, but \cite{solerKineticTomographyGalactic2025} used kinetic tomography to measure flows in a 1.2 kpc region around the Sun, finding converging mass inflow rates of a few solar masses per thousand years. This is more than what was found by \cite{beutherDynamicalCloudFormation2020}, who quantified the converging flows towards a dense zone within the G28.3 region, and found values around 0.05  solar masses per thousand years \citep[see also][for observations of converging flows at smaller scales]{beutherHierarchicalAccretionFlow2025}. It is still challenging to observationally determine what is driving these inflows.

In this paper, we investigate what fraction of mass inflow onto molecular clouds is gravitationally driven, meaning that it is driven by the combination of the self-gravity of the gas, and  the vertical gravitational potential in our simulations. The possible driving by the potential of stellar spiral arms is not considered, as this is not included in our simulations. Unlike previous studies that focused on instantaneous accelerations or energy considerations \citep[e.g.,][]{ibanez-mejiaGravityMagneticFields2022}, we integrate gas motions over 10 Myr, to capture the cumulative effect of gravity and average out random turbulence-driven fluctuations.

The structure of the paper is as follows. In Section~\ref{sec:method}, we describe the simulation and measurement methodology. Section~\ref{sec:results} presents results from a case study of a star-forming region. Limitations are discussed in Section~\ref{sec:caveats}, followed by a broader discussion in Section~\ref{sec:discussion}. Finally, Section~\ref{sec:concls} summarizes our findings and conclusions.

\begin{figure*}
    \centering
    \includegraphics[width=\linewidth]{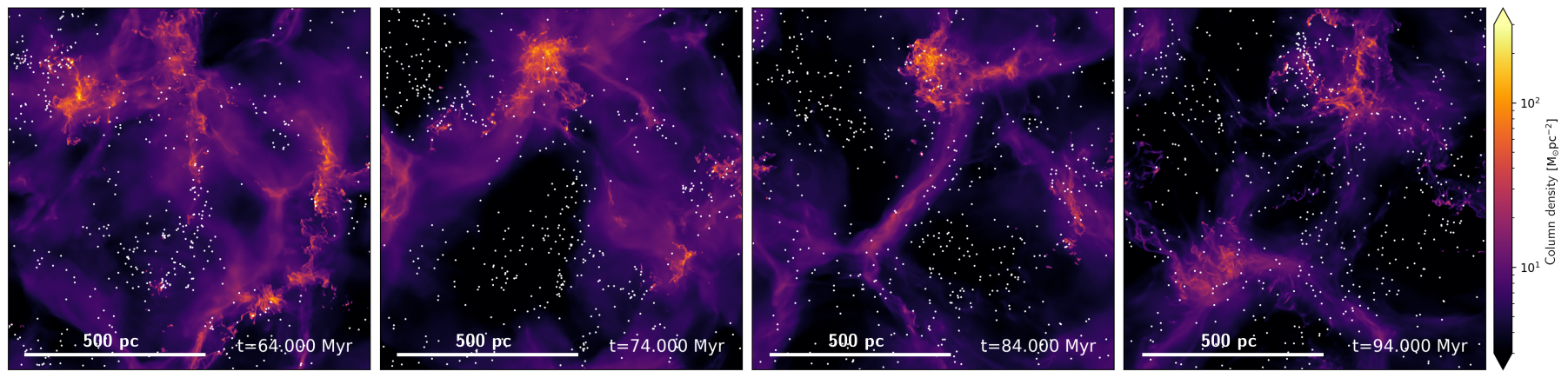}
    \caption{Evolution of the surface density of the full kpc$^3$ box from t = 64 Myr and t = 94 Myr. We study the timespan between  t = 74 Myr and t = 84 Myr, while the main over-dense structure already formed but is not destroyed by feedback yet. White dots represent the position of sink particles.}
    \label{fig:fullsim}
\end{figure*}

\begin{figure*}
    \centering
    \includegraphics[width=0.95\linewidth]{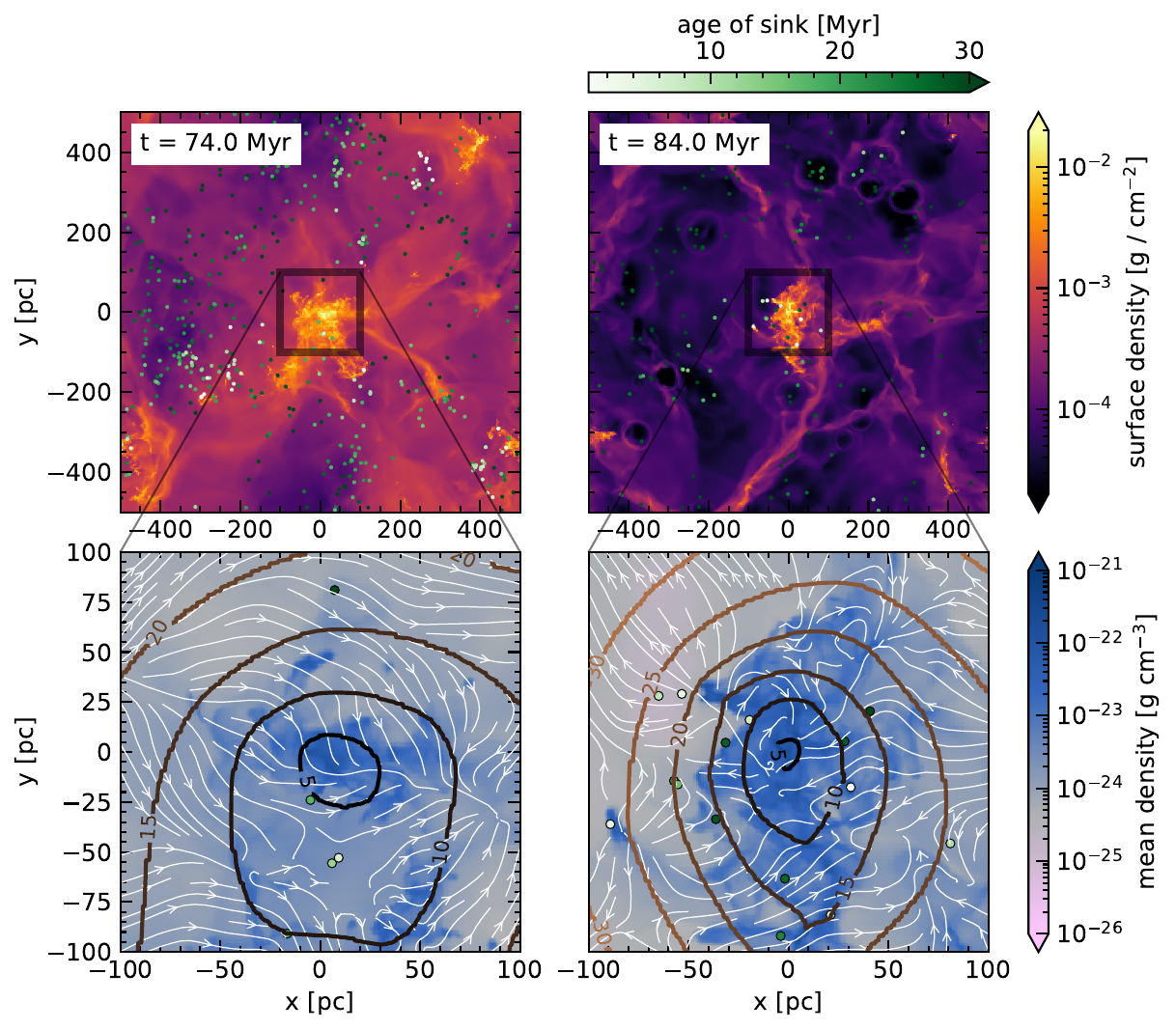}
    \caption{Overall face-on view of the simulation domain, centered on the main overdensity. Top: Surface density over a 100 pc vertical thickness at 74 Myr (left) and 84 Myr (right) of the stratified ISM box simulation, corresponding to the first and final snapshots we use for our case study. The map is centered around the minimum of the potential. The streamlines depict the direction of the velocity field. The sink particles (representing stellar clusters) are also depicted, with age coded in color.
    Bottom: Close-up view around the minimum of the gravitational potential for both snapshots. Here the mean density in a 20-pc thick slice is depicted. The contours lines represent the gravitational potential, in units of 10$^{7}$ cm$^2~$s$^{-2}$, and in a gauge where its minimum is set to zero.}
    \label{fig:situation}
\end{figure*}

\section{Method: gravitational contribution to the motion of tracers}
\label{sec:method}

This study quantifies the fraction of gas motion driven by the gravitational pull of gas overdensities as well as the gravity from the stars and dark matter, which are then compared with the convergent motions that are an integral part of the supersonic turbulence ubiquitously observed in the interstellar medium on galactic scales. We introduce gas tracers into a numerical ISM simulation and analyze the forces influencing their evolution over time. First, we describe the simulations used, then explain how we identify the dominant driver of tracer evolution.

\subsection{Stratified box simulation of the ISM} \label{subsec:simulations}

We use a stratified box simulation capturing the key physics of ISM dynamics at kiloparsec scales, including self-gravity, stellar disk potential, magnetic fields, cooling, star formation, and stellar feedback. Our goal is to assess whether the gas that feeds molecular clouds undergoes gravitational contraction, focusing on the relative importance of gravity compared to other forces such as pressure gradients and magnetic fields.

\subsubsection{General description of the simulations }

\begin{figure*}
    \includegraphics[width=\linewidth]{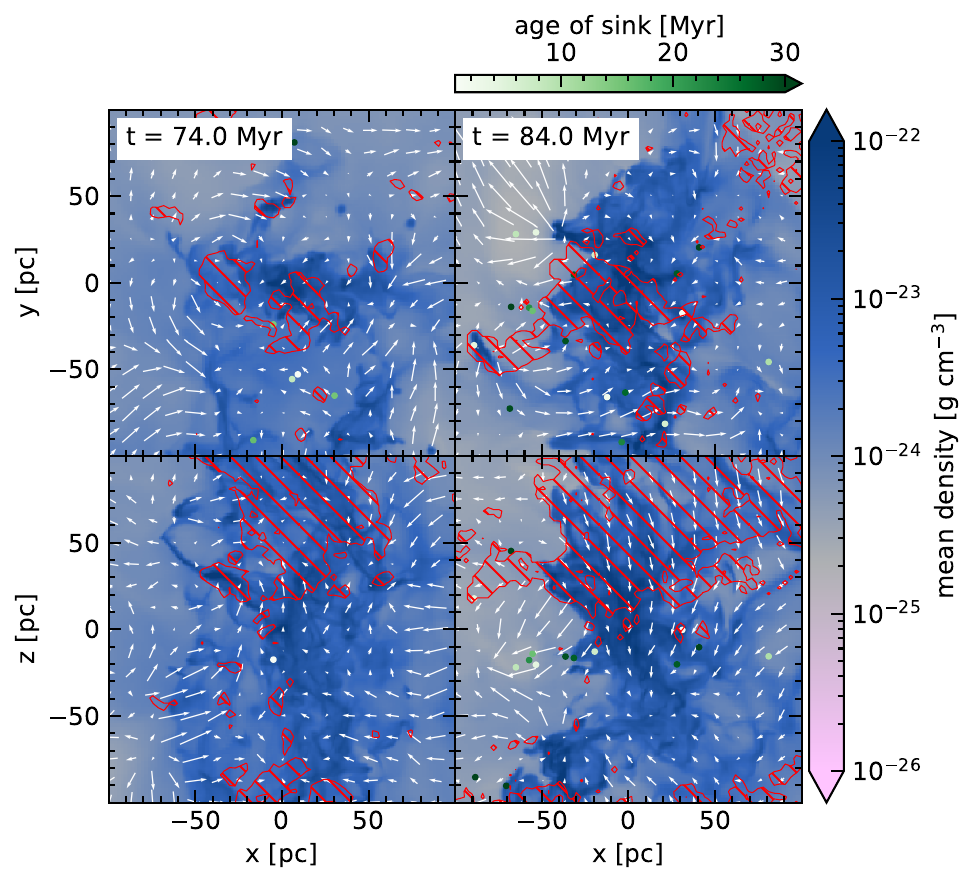}
    \centering
    \caption{Average density over a 20-pc thick slice around the minimum of the potential at the beginning (left) and the end (right) of the studied time-span. The dashed region shows where at least 20\% of the gas tracers are gravity-driven. The top panels show the face-on view while the bottom panels depict an edge-on view.}
    \label{fig:zoom_evo}
\end{figure*}

We re-simulate 10~Myr of the simulation \texttt{n0.66\_rms00000} presented by \citet[][Table~2]{colmanRoleTurbulenceSetting2025} using the magneto-hydrodynamics code \textsc{Ramses} \citep{teyssierCosmologicalHydrodynamicsAdaptive2002}. The setup has been developed and refined over the years (\citealt{iffrigStructureDistributionTurbulence2017,collingImpactGalacticShear2018,brucyLargescaleTurbulentDriving2020,colmanSignatureLargescaleTurbulence2022,brucyLargescaleTurbulentDriving2023}; for comparison see also \citealt{walchSILCCSImulatingLifeCycle2015}, \citealt{kimThreephaseInterstellarMedium2017} or \citealt{gurmanGHOSDTSimulationsMagnetic2025}). We model a 1~kpc$^3$ stratified, thermally bistable ISM region with adaptive resolution from 4 to 1~pc.
Periodic boundaries are applied in $x$ and $y$, and open boundaries in $z$. The simulation represents an environment which is slightly less dense than the Solar neighborhood. The interstellar gas is initially distributed as a Gaussian along the vertical axis, with a number density of hydrogen atoms.
\begin{equation}   \label{eq:n0} n(z) = n_0 \exp \left( - \frac{1}{2} \left ( \frac{z}{z_0} \right)^2 \right), \end{equation}
with $n_0 = 0.66\ \mathrm{cm}^{-3}$ and $z_0 = 150\ \mathrm{pc}$ (scale height), corresponding to a surface density of 8.4 $\Msun~\mathrm{pc}^{-2}$. 
In addition to the self-gravity of the gas, the gas also experiences an external gravitational potential from old stars and dark matter, given by
\begin{equation}
\label{eq:gext}
\vec{g}_{\mathrm{ext}}(z) =  - \left( \frac{a_1 z}{\sqrt{z^2+z_0^2}} + a_2 z \right) \vec{e}_z,
\end{equation}
with $a_1 = 1.42 \times 10^{-3}$ kpc~Myr$^{-2}$, $a_2 = 5.49 \times 10^{-4} \, \mathrm{Myr}^{-2}$, and $z_0 = 0.18$ kpc \citep{kuijkenMassDistributionGalactic1989, joungTurbulentStructureStratified2006}.
The simulation includes subgrid models for star formation (via sink particles) and stellar feedback (supernovae, uniform UV radiation proportional to the SFR, photoionization).
More precisely, a sink particle \citep{bleulerMoreRealisticSink2014} is created whenever the number density gets over $n_\mathrm{sink} = 10^3$~cm$^{-3}$ and represents a cluster of stars. It then accretes 10\% of the gas over $n_\mathrm{sink}$ within a radius of 4 pc at each timestep. Whenever the mass in the sink exceeds 120 $\Msun$, a stellar object, representing a massive star, is created within the sink. Its mass is randomly picked such that the initial mass function of massive stars follows a Salpeter slope \citep{salpeterLuminosityFunctionStellar1955}. The stellar object's ionizing radiation luminosity over its lifetime is deduced from its mass \citep{rosdahlRAMSESRTRadiationHydrodynamics2013,geenPhotoionizationFeedbackSelfgravitating2015,iffrigMutualInfluenceSupernovae2015}.

\subsubsection{Resimulation with tracer particles }

Figure~\ref{fig:fullsim} shows the face-on column density of the full simulation box from 64 to 94 Myr. We focus on a giant star-forming region, after the simulation has evolved sufficiently for at least one generation of stars to have formed and produced feedback. The region we study is a 100~pc-wide overdensity forming at $t \approx 64$~Myr and destroyed by stellar feedback ~20~Myr later\footnote{Typical lifetimes of observed clouds in nearby galaxies are similar \citep{chevanceLifecycleMolecularClouds2020}.}. It can be seen in the middle-top part of the column density maps of Figure \ref{fig:fullsim}. We specifically examine the 10 Myr timespan just before star formation starts within this cloud complex, re-simulating from $t_s = 74$~Myr until $t_e = 84$~Myr. 
The starting snapshot satisfies two criteria:
\begin{enumerate}
    \item SN-driven turbulence is well developed, with the first generation of massive stars having already exploded.
    \item It contains a clearly identifiable complex of clouds, which have not yet formed stars.
\end{enumerate}

We inject tracer particles to follow the gas flow, using the implementation of \cite{pichonRiggingDarkHaloes2011} and \cite{duboisFeedingCompactBulges2012}. One tracer is placed per cell, so that there are 63~705~314 tracer particles in total, with each one initially assigned the cell’s mass. Tracers follow the gas velocity computed with a cloud-in-cell scheme, assuming their volume equals that of the current cell. 

\subsection{Selection of gravity-driven inflowing gas}
\label{subsec:def}

To separate between gravity-driven and turbulence-driven gas motions, we use a procedure that takes into account the fact that the action of gravity tends to accumulate over time, while the effect of the other forces driving turbulence, in particular stellar feedback, is random and may partially cancel out over time\footnote{It is worth noting, however, that gravitational fragmentation causes new sites of collapse to appear over time, and therefore the gravitational acceleration on a tracer particle may also change in magnitude and/or direction. Therefore, this term may also be marginally affected by partial cancellations, albeit not as strong as those expected for random SN explosions.}. This point is further discussed in Appendix~\ref{sec:time_evo}. To this end, we compute the averaged gravitational acceleration over the studied timespan for each tracer $i$:
\begin{equation}
    \label{eq:a_grav}
    \left<\vec{a}_{\mathrm{grav},i}\right>(t_s, t_e) := \dfrac{1}{t_e -t_s} \int_{t_s}^{t_e} \vec{g_i}\ \mathrm{dt}.
\end{equation}

We also calculate the total averaged acceleration, including all contributions (pressure gradients, magnetic fields, supernova-driven turbulence), obtained from the change in velocity:
\begin{equation}
    \label{eq:a_tot}
    \left<\vec{a}_{\mathrm{tot},i}\right>(t_s, t_e)  := \dfrac{1}{t_e -t_s}  \int_{t_s}^{t_e} \vec{a_i}\ \mathrm{dt} = \dfrac{\vec{v_i}(t_e) - \vec{v_i}(t_s)}{t_e -t_s}.
\end{equation}
From Eqs~\eqref{eq:a_grav} and \eqref{eq:a_tot}, we can compute the acceleration from all the non-gravitational forces:
\begin{equation}
    \label{eq:other}
    \left<\vec{a}_{\mathrm{other},i}\right>(t_s, t_e)  :=   \left<\vec{a}_{\mathrm{tot},i}\right>(t_s, t_e) -  \left<\vec{a}_{\mathrm{grav},i}\right>(t_s, t_e).
\end{equation}

We then determine if gravity dominated the gas evolution by comparing the norms of $\left<\vec{a}_{\mathrm{grav},i}\right>(t_s, t_e)$ and $\left<\vec{a}_{\mathrm{other},i}\right>(t_s, t_e)$\footnote{We also tested an additional criterion requiring that the angle between $\left<a_{\mathrm{grav},i}\right>$ and the total acceleration  $\left<\vec{a}_{\mathrm{tot},i}\right>$ is less than 45\degree\ with no significant change to the results.}. A tracer is said to be gravity-driven if
\begin{equation}
\label{eq:norm}
\left<a_{\mathrm{grav},i}\right> > \left<a_{\mathrm{other},i}\right>.
\end{equation}
In the following, we quantify which fraction of the tracers contributing to the mass assembly of molecular clouds are gravity-driven. 

\begin{figure}
    \centering
    \includegraphics[width=\linewidth]{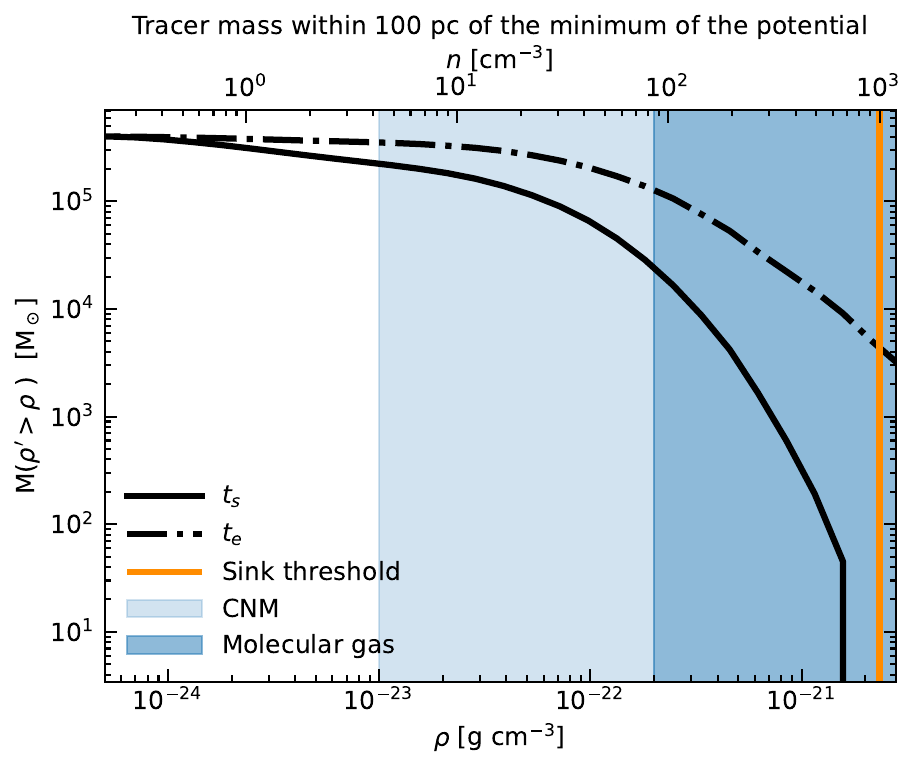}
    \caption{Enclosed mass within 100 pc of the minimum of the potential that is over a given density threshold at the starting time $t_s$ (solid line) and the end $t_e$ (dash-dotted) line. To help the interpretation, the sink formation threshold (vertical orange dotted line) is added, as well as the density thresholds above which the gas is expected to be in the CNM (light blue shaded area) or molecular (dark blue shaded area) phase components, based on \cite{drainePhysicsInterstellarIntergalactic2011}).}
    \label{fig:enclosed_mass}
\end{figure}

\section{Results}
\label{sec:results}

\subsection{General evolution of the simulation within the studied time-span}

Before addressing the role of gravity in the mass assembly of star-forming clouds, compared to all other processes at work, we first qualitatively describe the evolution of the selected cloud complex.

Figure~\ref{fig:situation}'s top panels show the column density of the simulation box, centered on the cloud complex, at the beginning and end of the studied timespan. The bottom panels show a thick density slice zoomed-in onto the cloud complex, with contours of the gravitational potential. All panels are centered around the minimum of the gravitational potential of the gas and the velocity vectors are plotted in its reference frame. 
At $t_s = 74$~Myr, the structure was initially surrounded by a significant amount of diffuse gas, and large voids containing a previous generation of stellar clusters, whose feedback drives strong diverging flows (left panels). Ten millions of years later, the density contrast with its direct environment increased dramatically, and it became more compact (right panels). Six new sink particles formed within the complex, two of which have already cleared their surroundings

Figure~\ref{fig:zoom_evo} provides a more detailed view with face-on and edge-on slices, highlighting regions where more the fraction of gravity-driven gas tracers is higher than the global average of 20\%  (red-hatched areas). It shows that of some of the gas assembly is gravity-driven, but not from all directions. Gas that ends up in the denser part of the clouds or that comes from the $z>0$ and $x>0$ directions seems more likely to be gravity-driven. Since the gravitational potential is very smooth (see Figure~\ref{fig:situation}), the asymmetry of the gravity-driven inflow is probably a consequence of local variation of the turbulence strength. This is line with the fact that the velocity (white arrows) within the red-hatched lines is relatively smaller than outside them.

Figure~\ref{fig:enclosed_mass} permits a more quantitative appreciation of the evolution of the cloud complex, by showing the total mass above a given density within 100~pc of the gravitational potential minimum of the gas. The total mass above $10^{-23}~\gpcc$ increases by 50\%, from $4\times10^5~\Msun$ to $6\times10^5~\Msun$, while the mass within molecular clouds grows fivefold, from $4 \times 10^{4}~\Msun$ at $t_s$ to $2 \times 10^{5}~\Msun$ at $t_e$.

 \begin{figure}
     \centering
     \includegraphics[width=\linewidth]{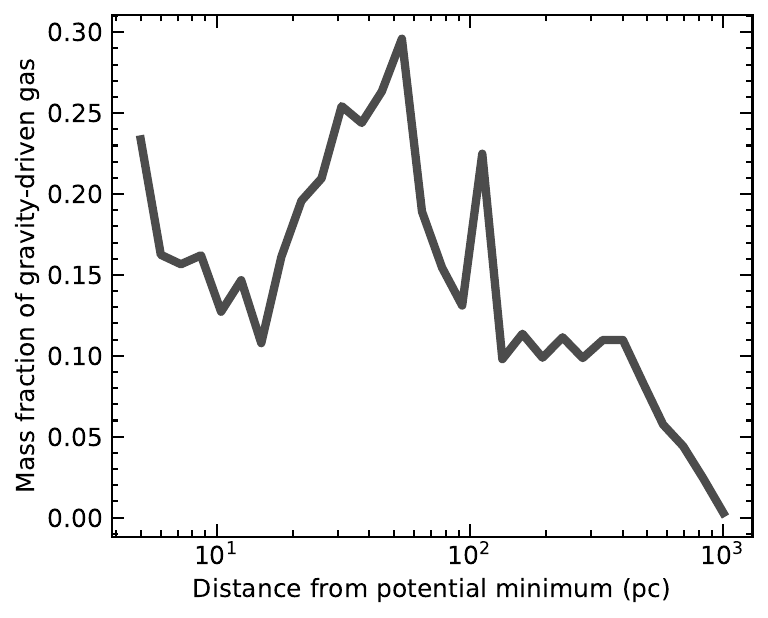}
    \caption{Fraction of the total mass in tracer particles in gravity-driven particles during the 10-Myr time span, as a function of distance from the minimum of the potential.}
     \label{fig:mass_fraction}
 \end{figure}

\begin{figure}
    \centering
    \includegraphics[width=\linewidth]{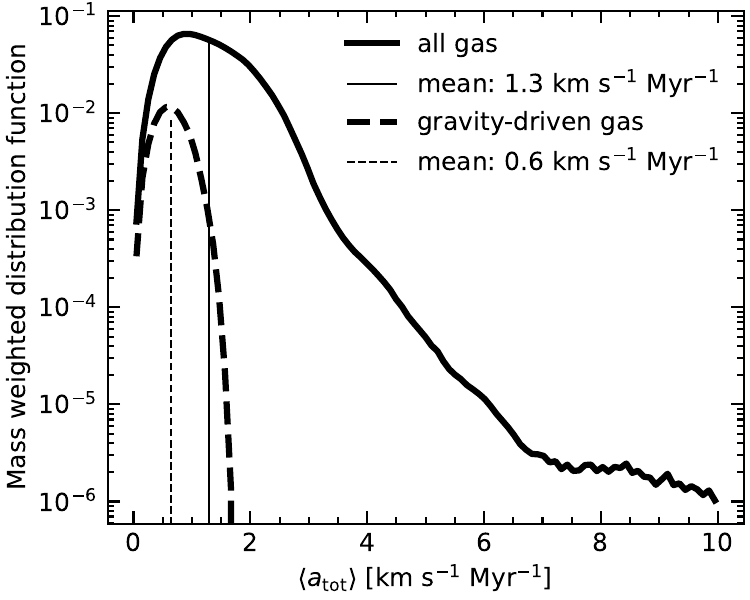}
    \caption{Mass-weighted distribution of the averaged total acceleration $a_\mathrm{tot}$ (solid line) of the gas tracers. The contribution of gravity-driven gas tracers is indicated by the dashed line. The vertical lines show the mass-weighted averaged of both distributions.}
    \label{fig:distrib_vdiff}
\end{figure}

\subsection{Effective range of gravity}
\label{subsec:mass_fraction}

We quantify the fraction of gas around the cloud complex contributing to cloud mass assembly that is gravity-driven (Section~\ref{subsec:def}). Figure~\ref{fig:mass_fraction} shows the mass fraction of gravity-driven tracers relative to the total, as a function of distance from the potential minimum at $t_e$.
Up to 100~pc from the potential minimum, 10--30\% of tracers are gravity-driven, a modest but clearly noticeable fraction. We note that this fraction decreases at larger distances.

The actual contribution of gravity-driven gas to cloud mass assembly also depends on inflow velocity. At 100~pc scales, gravitational acceleration is typically much smaller than other forces. In Figure~\ref{fig:distrib_vdiff} we present the mass-weighted distribution of total acceleration $a_\mathrm{tot}$ and the subset of gravity-driven tracers. The mean acceleration of gravity-driven gas is roughly half the global mean, below 2~$\kms~\mathrm{Myr}^{-1}$, while total acceleration can reach several tens of $\kms~\mathrm{Myr}^{-1}$. Evaluating the mass inflow contribution of gravity-driven gas must account for this velocity difference, which is addressed in the next section.

\subsection{Hierarchical mass inflow}
\label{subsec:HMI}

\begin{figure*}
    \centering
    \includegraphics[height=0.29\textheight]{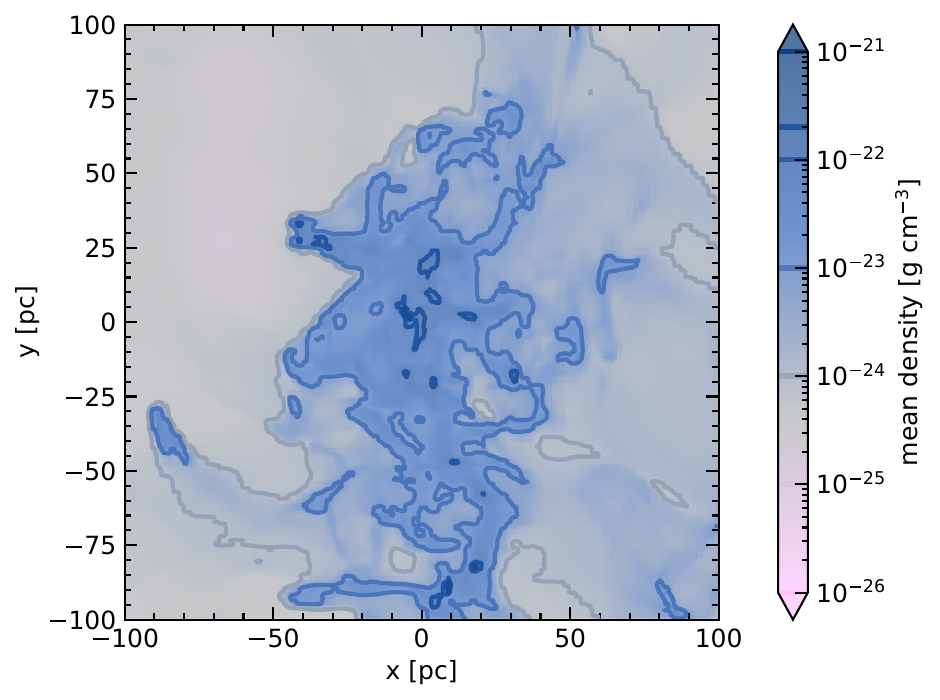}
    \includegraphics[height=0.29\textheight]{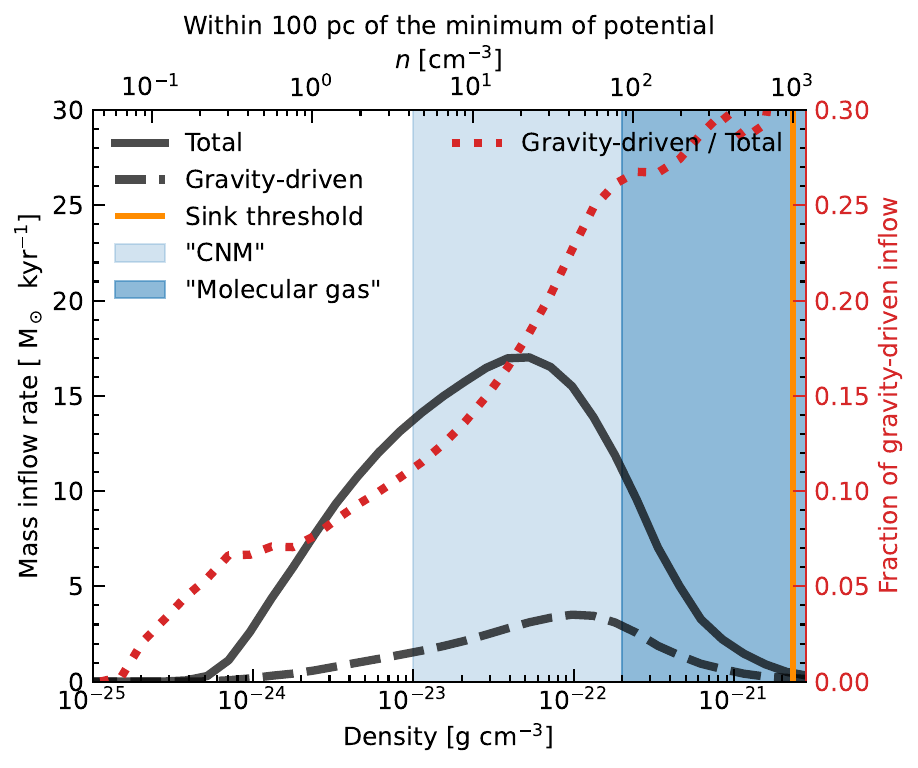}
    \caption{Left: 20pc-thick slice of the density at $t_e$, with density contours, illustrating the hierarchical nature of the cloud complex. Right: Mass inflow across density layers according to Eq.~\eqref{eq:mass-inflow} for all the tracers that end up within 100 pc of the minimum of the potential (black solid line) and for gravity-driven tracers (black dashed line). As a reading help, the sink formation threshold (vertical orange dotted line) as well as the density thresholds over which the gas is expected to be in the the CNM (light blue shaded area) or molecular (dark blue shaded area) phases are also added (based on \cite{drainePhysicsInterstellarIntergalactic2011}).
    The ratio between the total flow and the gravity-driven flow (red dotted line) can be read on the right vertical axis.}
    \label{fig:mass_inflow_density}

\end{figure*}

In the GHC scenario, gas accretes hierarchically to increasingly dense regions, forming a hierarchy of fragmented structures \citep{vazquez-semadeniGlobalHierarchicalCollapse2019}. There is no single collapse point; so testing GHC requires analyzing the force driving inflows across density layers.
Following this logic,  we compute the mass flow from lower to higher density regions:

\begin{equation}
\label{eq:mass-inflow}
 \dot{\mathrm{M}}_\mathrm{H}(\rho, t_\mathrm{s}, t_\mathrm{e}) = \dfrac{1}{t_e - t_s}\sum_{i,\ \rho_i(t_s)<\rho < \rho_i(t_e)} m_i,
\end{equation}
with $\rho_i$ the density of the cell the tracer $i$ is currently in. In words, this equation says that the hierarchical mass inflow rate $\dot{\mathrm{M}}_\mathrm{H}(\rho, t_\mathrm{s}, t_\mathrm{e})$  across a given density $\rho$ between the starting time $t_s$ and the end time $t_e$ is the sum of the mass of the tracers particles that belonged to a cell with a density lower that $\rho$ at $t_s$ and ended up in a cell with a density higher than $\rho$ at $t_e$, divided by the integration time $t_e - t_s$. One can then define the gravity-driven hierarchical inflow rate, by further selecting the tracers that are gravity driven between $t_s$ and $t_e$ in the summation.

Right panel of Figure~\ref{fig:mass_inflow_density} shows the hierarchical inflow rate across density layers. Only a small fraction of the total inflow is gravity-driven: roughly 10\% of tracers reaching densities above $10^{-23}\ \gpcc$ (the expected transition point from the warm to the cold ISM phase) are gravity-driven. This fraction increases for denser thresholds: $\sim$20\% of tracers reaching $10^{-22}\ \gpcc$ are gravity-driven. Above this threshold, most gas is expected to be molecular. Within molecular clouds, the fraction of gravity-driven inflow grows with density. In our simulation, densities where over 50\% of the inflow is gravity-dominated are not reached, suggesting overall gravitational collapse occurs at densities higher than our sink formation threshold, $2.34\times10^{-21}\ \gpcc$, in line with observational estimates \cite[e.g.,][]{Lada2010, traficanteMultiscaleDynamicsStarforming2020} and theoretical estimates \cite[e.g.,][]{Krumholz2009, Clark2014}.

It is known that velocity-advected tracers are systematically biased toward converging flows \citep{priceComparisonGridParticle2010,genelFollowingFlowTracer2013}; thus, absolute flow rates in Figure~\ref{fig:mass_inflow_density} could be overestimated. In Section~\ref{subsec:limit_tracer} we show that in our simulation, the drift between the gas and the tracers is limited.

 \subsection{Gas self-gravity and external potential}

In our simulation, the gas is subject to its own gravity as well as the one from the stars and dark matter, the latter modeled by a vertical analytical potential (Eq.\ \ref{eq:gext}).  In Figure~\ref{fig:zoom_evo}, we see that gravity-driven gas inflowing onto the cloud comes from both the horizontal and vertical direction, indicating that both may play a role. 

The integrated gravitational acceleration $\left<\vec{a}_{\mathrm{grav},i}\right>(t_s, t_e)$ includes both of these contributions.
To isolate the effect of self-gravity, we compute $\left<\vec{a}_{\mathrm{SG},i}\right>(t_s, t_e)$ defined by
\begin{equation}
\label{eq:asg}
     \left<\vec{a}_{\mathrm{SG},i}\right>(t_s, t_e) = \left<\vec{a}_{\mathrm{grav},i}\right>(t_s, t_e) - \int_{t_s}^{t_e} g_\mathrm{ext}(z_i(t))\mathrm{d}t
\end{equation}
where $z_i(t)$ is the vertical position of the tracer $i$ at time $t$.
In practice, since the positions of the 60 million tracers are stored every Myr, a linear interpolation of the $z$ coordinate is used to compute \eqref{eq:asg}.
We can then determine the tracers that are self-gravity-driven, by using the criterion given by Eq.~\eqref{eq:norm} substituting  $\left<\vec{a}_{\mathrm{grav},i}\right>(t_s, t_e)$ by $\left<\vec{a}_{\mathrm{SG},i}\right>(t_s, t_e)$.

\begin{figure}
     \centering
     \includegraphics[width=0.93\linewidth]{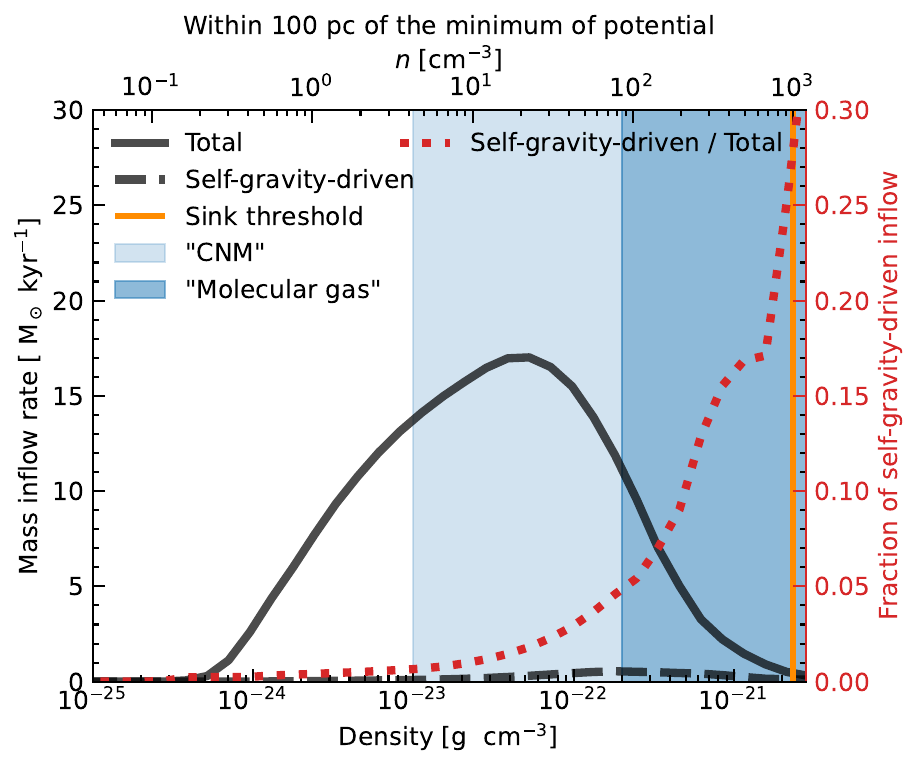}
     \caption{Same as Figure~\ref{fig:mass_inflow_density} (right), but  with a criterion selecting only the self-gravity-driven gas (Eq.~\ref{eq:asg}).}
     \label{fig:mass_inflow_sg}
 \end{figure}

Figure \ref{fig:mass_inflow_sg} shows the mass inflow across density layers with this updated criterion. Adopting the same density thresholds for phase definition as in Section \ref{subsec:HMI}, we note that the self-gravity-driven flow onto the CNM is negligible and that the self-gravity-driven flow onto the  molecular clouds is about 3\%.
However, inside molecular clouds, the fraction of the self-gravity-driven mass inflow is increasing very sharply.
This points towards a change of the nature of the inflow responsible of the increase of the density, from mostly driven by compressible turbulence with a small contribution from the large-scale external gravitational potential to mostly driven by the self-gravity of the gas. We discuss this transition in more details in section ~\ref{subsec:transition}.

\section{Caveats}
\label{sec:caveats}

\subsection{How representative is our model?}

The work we present here is a case study over one large simulated cloud complex in a numerical simulation. It encompasses several star forming clouds (see left panel of Figure~\ref{fig:mass_inflow_density}).
Using the ECOGAL wrapper \citep{colmanCloudPropertiesSpatial2024} around the HOP cloud detection algorithm \citep{eisensteinHOPNewGroupFinding1998}\footnote{with a threshold of $5\times 10^{-22} \gpcc$, a peak factor of 2, a saddle factor of 1 and a minimum number of 100 cells}, we identify 10 clouds in the cloud complex at the beginning of the integration time, and 33 at the end (see Appendix~\ref{sec: clouds} and Figure~\ref{fig:clouds_pos}). Their sizes vary from 6 to 30 pc, their masses from $5\times 10^2$ to $10^4\ \Msun$, and they have velocity dispersions from 1 to 4 $\kms$ with typical values for clouds of these sizes (see Figure~\ref{fig:clouds_rel}).
While this ensures that various cloud properties are probed, it cannot be generalized to all possible star-formation environments. A possible follow-up would be to determine how the fraction of gravity-driven mass inflow onto molecular clouds depends on broader environmental conditions, such as the total mass of the star forming complex or its velocity dispersion. At smaller scale, looking at how it depends on the properties of individual clouds is also an important pending prospect.

\subsection{Limitations of tracer particles}
\label{subsec:limit_tracer}

Velocity-advected tracers, like those used in this work, are known to poorly reproduce the gas density field over long integration times \citep{priceComparisonGridParticle2010,genelFollowingFlowTracer2013}, since they tend to accumulate in cells with converging flows.
\textsc{Ramses} offers a more accurate Monte-Carlo tracer particle implementation \citep{cadiouAccurateTracerParticles2019}, but it is not suitable for the analysis performed in this work. In the Monte-Carlo approach, tracers jump between cells, so their velocity does not directly reflect the forces acting on the gas.
Therefore, we employ velocity-advected tracers and limit their inaccuracy by focusing on a single star-formation cycle, reducing cumulative drift from multiple converging and diverging flows.
We quantify the error by comparing the gas density in the mesh to the one reconstructed from the tracer positions and masses through a cloud-in-cell algorithm (Figure~\ref{fig:tracer_err} in Appendix~\ref{sec:tracer_drift}).
For over 70\% of the mass, the reconstruction error is below 15\%, which is acceptable for this study, since our focus is on the ratio of gravity-driven to total inflow rather than the absolute flow values.

\subsection{Resolution}
\label{subsec:resolution}

The maximal resolution of our simulation is 1 pc, and the the maximum density we can self-consistently reach is capped by the sink formation threshold of $10^3$ cm$^{-3}$. While this is sufficient to study the mass assembly of the cloud, which is the primary goal of this study, its does not meet the requirement to accurately model the internal dynamics of the clouds \citep[see, e.g.,][]{seifriedSILCCZoomDynamicChemical2017}. As such, it is worth recalling that we do not discuss the instantaneous balance of forces of the gas inside the clouds, but rather the driving force responsible for its velocity, which reflects the balance of force during its past evolution.

\subsection{Behaviour near the sink particles}
\label{subsec:tracer_sink}

Sink particles progressively absorb the gas above the sink formation threshold of $10^3$~cm$^{-3}$, within an accretion radius of 4 pc. As a consequence, the balance of forces that apply to the gas within in vicinity of the sinks may be altered, with a sudden reduction of the thermal and ram pressure forces. 
However, this effect is typically negligible due to the rather small volume of the sink accretion zones, and the small number of sinks within the studied complex of clouds. Indeed, in the last snapshot, over the full simulation only 0.11\% tracers are within the accretion radius of a sink, and 0.58\% within twice the accretion radius of a sink.
In Appendix \ref{sec:tracers_sink}, we further study the impact of the effect of the accretion zone on our results by recomputing the fraction of gravity-driven and self-gravity-driven mass inflows (from respectively Figure~\ref{fig:mass_inflow_density} and Figure~\ref{fig:mass_inflow_sg}), this time excluding all the tracers that end up within two accretion radii from a sink. The results are almost unchanged, with only slight variations that we discuss in Appendix \ref{sec:tracers_sink}.

\subsection{Neglect of spiral arm gravitational potential} \label{sec:spiral_pot}

One important dynamical ingredient not included in our simulations is the gravitational potential of stellar spiral arms.
Nevertheless, this may be an important gas-gathering mechanism, 
since it is clear that, at least in grand-design spiral galaxies, the largest  molecular gas complexes roughly coincide with the stellar spiral arms. In this case, the large-scale molecular gas assembly mechanism could be the gravity of the stellar arms, with SN driving mainly stirring and rearranging this gas, and assembling new GMCs as large concentrations within this mostly molecular large-scale environment. This, however, is beyond the scope of the present study.

\section{Discussion} \label{sec:discussion}

\subsection{Implications for analytical models for the star formation rate}

 There is a renewed interest in developing and improving analytical prescriptions for the SFR \citep{zamora-avilesEvolutionaryModelCollapsing2012,zamora-avilesEvolutionaryModelCollapsing2014, burkhartStarFormationRate2018,ostrikerPressureregulatedFeedbackmodulatedStar2022,hennebelleInefficientStarFormation2024,brucyInefficientStarFormation2024,meidtReconcilingExtragalacticStar2025}, which are useful as subgrid models in large-scale numerical simulations, to interpret observations or simply to better understand the scenario of star formation.
 These models share the idea that the star formation rate can be derived from the distribution function of the density by integrating the mass of gas in gravitationally unstable structures or above a given density threshold, and dividing by the time needed for the gas to collapse. A review of the  models based on turbulent support as well as their shortcomings is given by \cite{hennebelleInefficientStarFormation2024}, while a review of the models based on cloud-scale gravitational contraction is given in \citet{vazquez-semadeniGlobalHierarchicalCollapse2019}.
 
 In the context of the turbulent support models \citep{krumholzFormationStarsGravitational2005,padoanStarFormationRate2011,hennebelleAnalyticalStarFormation2011,Hopkins2012,federrathStarFormationRate2012}, the underlying concept is that interstellar turbulence is shaping the density distribution of the ISM, with only the densest regions undergoing gravitational collapse and forming stars. As a consequence, the density distribution produced by turbulence, usually assumed to be a lognormal \citep{vazquez-semadeniHierarchicalStructureNearly1994,nordlundDensityPDFsSupersonic1999,federrathDensityProbabilityDistribution2008}, is used.
 However, it is now established from observations \citep{kainulainenProbingEvolutionMolecular2009, Lombardi2014, Schneider2015a, Schneider2015b, Veltchev2019, Pezzuto2023} as well as simulations and theoretical considerations \citep{klessenOnePointProbabilityDistribution2000, federrathStarFormationEfficiency2013,kritsukDensityDistributionStarforming2011,girichidisEvolutionDensityProbability2014,leeStellarMassSpectrum2018,khullarDensityStructureSupersonic2021,Marinkova2021, Veltchev2024, Donkov2025} that self-gravitating systems will develop a power-law tail in the high density regime as the gas collapses. It has been argued that this power law  \cite[{e.g.,}][]{burkhartStarFormationRate2018} should be taken into account for the computation of the SFR.
 This is particularly relevant in globally collapsing systems, when estimating the SFR over timescales shorter than the overall collapse time of the system.
 Indeed, our present study supports the idea, mentioned by \cite{hennebelleInefficientStarFormation2024}, that whether to take the power-law tail into account or not depends on the spatial scales and timescales considered.
 Gravity-driven motions play a small but non-negligible role in the formation and mass assembly of large molecular clouds (more than a few parsecs). As such, large regions of several hundred parsecs will not be in global collapse and models relying on the lognormal density distribution may be more relevant. On the other hand, we show that the ratio of gravity-driven inflow quickly increases inside molecular clouds, meaning that taking into account the density power law due to gravity is crucial when studying the SFR of an individual cloud.

In the GHC model,
the cloud-scale contraction provides an evolutionary framework for molecular clouds and their star formation activity (the SFR and the range of stellar masses present in the clouds).
The clouds grow in mass by accretion from their environment. They fragment through a Hoyle-like \citep{Hoyle53} gravitational fragmentation process, which is seeded by turbulence-induced {\it nonlinear} density fluctuations.
As the clouds grow, they simultaneously increase their SFR, and form stars with a mass distribution extending to increasingly larger masses.
This process continues until the newly formed massive stars begin to disrupt the clouds with pre-SN feedback, thus decreasing the clouds' masses and SFR alike \citep{zamora-avilesEvolutionaryModelCollapsing2012,zamora-avilesEvolutionaryModelCollapsing2014, vazquez-semadeniGlobalHierarchicalCollapse2019}. In this context, our results that the velocities in the density range sampled by our simulation are mostly the result of the external compressions rather than from the cloud's self-gravity imply that the cloud assembly timescale is also determined by the cumulative effect of the external compressions. And since these velocities can be larger than the gravitational velocity (see Figure \ref{fig:linewidth}), the cloud assembly timescale may be even shorter than that expected from self-gravity.

\subsection{Contribution of the gravity-driven gas to the linewidth}
\label{subsec:linewidth}

\begin{figure}
    \centering
    \includegraphics[width=\linewidth]{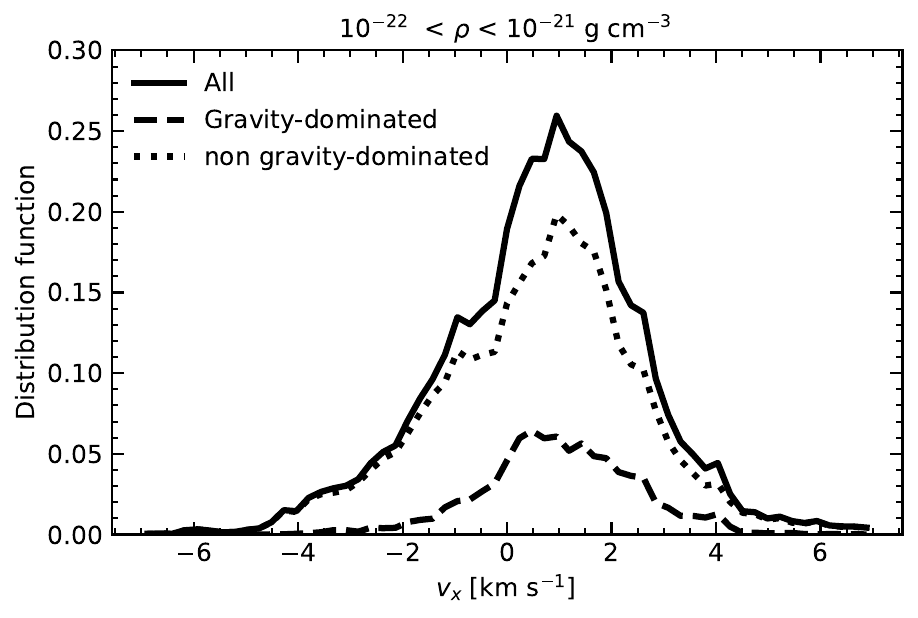}
    \caption{Mass-weighted distribution function of velocities of the dense gas ($10^{-22}  < \rho < 10^{-21}  \gpcc$) along the $x$ direction of the gas at $t=t_e$, (solid line), within a 100 pc$^3$ around the minimum of the potential. The contribution of gravity-driven gas is shown (dashed line). The velocities are in the rest-frame of the minimum of the potential.
    The mass of gravity-driven gas is estimated by multiplying the mass of the cell by the fraction of gravity-driven tracers in each cell.}
    \label{fig:linewidth}
\end{figure}

A key question within the debate between turbulence driven models of MCs and the GHC is how to interpret the supersonic linewidths of molecular tracers. In the first scenario, they are interpreted as the signpost of turbulent motion, which itself is sometimes interpreted as providing support. Instead, in the latter they are thought to contain an important contribution from gravitational infall motions alongside large-scale compressible turbulent motions that promote collapse rather than opposing it (see Section \ref{sec:intro}). 
Without doing a complete synthetic observation, we can estimate the contribution of gravity-driven motion to the linewidth by using the mass-weighted distribution function of the velocity in a given line-of-sight.
Figure~\ref{fig:linewidth} shows that the distribution of the velocities of the gravity-driven gas is narrower than the full distribution (in the $x$ direction). In this figure, we consider only the gas with densities between $10^{-22} \gpcc$ and $10^{-21} \gpcc$ within a $(100~{\rm pc})^3$ box around the minimum of the potential. The contribution of gravity-driven gas is computed by multiplying the mass in each given cell by the fraction $f_g$ of gravity-driven tracers in that cells.
The full-width-half-maximum are respectively 3.8, 2.8 and 3.8 $\kms$ for the total linewidth, the hypothetical gravity-dominated linewidth and non-gravity-dominated linewidth.
We can also compare the velocity dispersion associated with the dense gas $\sigma_{x, \mathrm{d}}$ to the velocity dispersion of the gravity-driven dense gas, $\sigma_{x, \mathrm{g, d}}$ defined by
\begin{align}
    m_\mathrm{d} &=  \sum_{\rho_j > 10^{-22} \gpcc} m_j ,\\
    m_\mathrm{g,d} &=  \sum_{\rho_j > 10^{-22} \gpcc} f_g m_j ,\\
    \overline{v_{x,\mathrm{d}}} &= \sum_{j, \rho_j > 10^{-22} \gpcc} m_jv_{j,x} \\
    \sigma_{x, \mathrm{d}} &= \sqrt{ \dfrac{1}{m_\mathrm{d}} \sum_{j, \rho_j > 10^{-22} \gpcc} m_j \left(v_{j,x} - \overline{v_{x,\mathrm{d}}}\right)^2}  ,\\
     \sigma_{x, \mathrm{g, d}} &=  \sqrt{ \dfrac{1}{m_\mathrm{g, d} }\sum_{j, \rho_j > 10^{-22} \gpcc} f_g m_j \left(v_{j,x}-\overline{v_{x,\mathrm{d}}}\right)^2},
\end{align}
where $\rho_j$,  $m_j$, and $v_{j,x}$ are respectively the density, the mass and the velocity in the $x$ direction of the cell $j$.
The velocity dispersion of the dense gas in the 100 pc$^3$ box around the minima of the potential is $\sigma_{x, \mathrm{d}} = 1.9\ \kms$ while the velocity dispersion of the gravity-driven gas is smaller $\sigma_{x, \mathrm{g,d}} = 1.5\  \kms$.
The contribution of the dense gravity-driven gas on the total variance of the dense gas velocity is given by 
\begin{equation}
    \dfrac{ \sigma_{x, \mathrm{g, d}}^2}{\sigma_{x, \mathrm{d}}^2} \dfrac{m_\mathrm{g,d}}{ m_\mathrm{d}} \approx 0.1.
\end{equation}
We conclude that the contribution of gravity-driven gas to the linewidth of the gas at densities between $10^{-22} \gpcc$ and $10^{-21} \gpcc$ and at the 100~pc scale is negligible. 

\subsection{Transition from turbulence to the gravity-dominated regime}
\label{subsec:transition}

Our analysis considers the cumulative effect of both the total and the gravity-driven motions in the assembly of the cloud over a period of 10 Myr, and shows that, up to the densities and size scales that our simulation can resolve, it is essentially performed by the feedback-driven turbulent motions. However, this very process causes a growth of the cloud's mass (Figure \ref{fig:enclosed_mass}), and therefore, of its gravitational binding, thus increasing the contribution of gravity-driven motions in the ``molecular gas'' density regime. Indeed, Figure \ref{eq:mass-inflow} shows a steady increase in the contribution of gravity to the mass flow within molecular clouds as the density increases, and, at face value, a linear extrapolation of the fraction of gravity-driven inflow as a function of the logarithmic density (red dotted line) suggests that gravity could become the dominant force bringing gas to densities between $10^{-20}$ and $10^{-19} \gpcc$ during this period. These estimates are not impacted by the behavior of the gas close to the sinks, as discussed in Section~\ref{subsec:tracer_sink}.

Moreover, the effect of self-gravity in the assembly of the structures in the molecular gas density regime must increase over time precisely as a consequence of the mass accumulation there. 
This would increase the gravitational acceleration generated by this material. Indeed, a temporal plot of the gravity-driven inflow (right panel of Figure \ref{fig:time_evo} in Appendix~\ref{sec:time_evo}) shows that it increases monotonically as larger values of the ending time $t_e$ are considered for the calculation of the averaged accelerations in equations \eqref{eq:a_grav} and \eqref{eq:a_tot}, reaching almost half of the total when $t_e = 79$ Myr. Larger ending times give smaller gravity-driven fractions inside the molecular clouds because of the formation of sink particles that inject photoionizing feedback, and thus begin to disperse the dense gas. That is, the relative importance of gravity is increasing over time precisely due to the initially turbulence-dominated mass assembly.

This mechanism may explain why molecular clouds generally appear to be close to equipartition between their kinetic and gravitational energies \citep[e.g.,] [] {larsonTurbulenceStarFormation1981, heyerReExaminingLarsonsScaling2009, miville-deschenesPhysicalPropertiesMolecular2017}: rather than providing support against gravity, the large-scale, external SN-driven motions are assembling the clouds, thus increasing their gravitational energy, until it becomes comparable to the turbulent cloud-assembling motions. At that point, gravitational contraction may form the internal substructures of the molecular clouds (hub-filament systems, cores), giving them their characteristic flow patterns \citep[e.g.,] [] {GV14}, and culminating into stars, which then produce photoionizing feedback that disrupts the clouds. This is suggested by the fact that the gravity-driven motions are seen to decrease again after $t = 80$ Myr in the right panel of Figure \ref{fig:time_evo}, when this type of feedback turns on in the interior of the cloud in the simulation.

\subsection{Comparison with previous work}

In Section \ref{subsec:transition}, we discuss the fact that gravity could become the dominant force bringing gas to densities between $10^{-20}$ and $10^{-19} \gpcc$ over a period of a few million years. 
When comparing with previous studies, we have to keep in mind this statement applies to the past evolution of the gas inside the clouds, as it was getting denser. 
This is formally different than looking at the dominant force or energy within the cloud at a given time, even if the two are linked. We also have to keep in mind that we do not resolve the interior of the clouds (see Section \ref{subsec:resolution}), contrary to other studies that focuses on smaller scales.

On the observational side, \cite{traficanteMultiscaleDynamicsStarforming2020} proposes a critical column density for the transition into the gravity-dominated regime of $\Sigma \approx 0.1\ \mathrm{g}\ \mathrm{cm}^{-2}$. This threshold, assuming a typical size of 1 pc, corresponds to a density of the order of $10^{-20} \gpcc$ \cite[see also,][]{Lada2010}. This is compatible with our estimate for the transition into the gravity-dominated regime.

Our estimate is also compatible with numerical studies that simulate these scales but focuses on a instant comparision of the forces. By running a series of simulations of an isolated molecular cloud with different physical processes included, \cite{appelWhatSetsStar2023} find that above $5 \times 10^{-19} \gpcc$, the behavior of the compressing gas in the run including turbulence driving, magnetic field and jets is similar to the run including only gravity, suggesting that the accretion at these densities is mostly due to gravity. This regime would correspond to a state in which gravity account for almost 100\% of the flow, which is much stronger than saying that gravity is the main force driving the flow. As such, it is not suprising that it occurs at even higher densities.
A key marker of the dominance of gravity is the development of a power-law tail in the density distribution function of the cold gas \cite[e.g.,][]{girichidisEvolutionDensityProbability2014,leeStellarMassSpectrum2018, Veltchev2024}. Such a power-law is not clearly seen in our simulation (see Figure~\ref{fig:pdf_200pc} in Appendix~\ref{sec:pdf}), which is consistent with the fact the most of the flow is not gravity-driven.

Another study, by \cite{ibanez-mejiaGravityMagneticFields2022}, cover a wide range of scales including the ones we study here, by re-simulating in a selection of gravitationally unstable clouds extracted from a kpc-sized simulations. Turbulence in their simulation is driven by randomly injected supernovae and initially run without self-gravity, which is activated when they start their analysis.
In their Figure 4, they plot the mean accelerations of the gas from gravitational potential gradients, pressure gradients and magnetic field.  This approach allows one to determine what is the instantaneous dominant force at a given time. It is not suitable to estimate the contribution of each force on the mass inflow, which requires to integrate the accelerations over time. It is interesting to note that the density at which the gravitational acceleration evolves over times is different from one cloud to another. It is included in a range between $10^{-21}$ and $10^{-20}\ \gpcc$, which is consistent with our study.

\subsection{The roles of the SN and pre-SN feedback} \label{sec:pre-SN_vs_SN}

Our results suggest that a crucial role of the SN feedback is to assemble the clouds, while that of the pre-SN feedback is to disperse them.  Indeed, Figure \ref{fig:fullsim} and an inspection of an animation of the simulation\footnote{available here: \url{https://noe.brucy.fr/images/mass_inflow_column_density_z.webm}} shows that the cloud we analyze is formed by a compression due to the explosion of a previous generation of SNe, while Figure \ref{fig:time_evo} in Appendix \ref{sec:time_evo} shows that the reduction in gravity-driven velocities occurs at $t \sim 80$ Myr, when a couple of sink particles form in the cloud and begin to feed back on it via photoionizing radiation. This happens because the massive stars produce photoionizing radiation immediately after they form, and thus apply it in the interiors of the clouds, while SN explosions occur after the gas has been cleared out by the photoionizing radiation, and thus it can further compress the gas.

\subsection{Additional sources of turbulence}

In the simulation, turbulence is self-consistently driven by stellar feedback, particularly supernovae. This is in line with the finding that feedback is the main source of turbulence in Milky Way–like galaxies
\citep{bacchiniEvidenceSupernovaFeedback2020}.
However, this is likely not the case in denser, more gas-rich environments, where turbulence generated by galactic-scale motions is thought to contribute significantly to driving turbulence at the kiloparsec scale \citep{klessenAccretiondrivenTurbulenceUniversal2010,krumholzTurbulenceInterstellarMedium2016,krumholzUnifiedModelGalactic2018,brucyLargescaleTurbulentDriving2020,ejdetjarnGiantClumpsClouds2022}.
One might think that an increase in turbulence strength would reduce the role of gravity. However, turbulence is known to also produce large density contrasts, which can trigger gravitational collapse \cite[e.g.,][]{maclowControlStarFormation2004}, and this is  especially true when the driving source of turbulence excites compressive modes \citep[e.g.,][]{federrathComparingStatisticsInterstellar2010,menonCompressiveNatureTurbulence2021,brucyLargescaleTurbulentDriving2023}. Ultimately, the same experiment could be conducted in a similar ISM box with additional external large-scale driven turbulence (like the one used in \citealt{brucyLargescaleTurbulentDriving2023}), but this is beyond the scope of the present paper.

\newpage

\section{Conclusions} \label{sec:concls}

In this work, we quantify the contribution of gravity to the mass assembly of clouds and the flow of gas within them, and we compare it to the contribution from turbulent-driven convergent flows. Using a stratified box MHD simulation of the ISM that includes SN and photo-ionization feedback, we consider scales from 100 pc to 1 pc, with densities ranging from $10^{-25}$ to $10^{-21}\ \gpcc$.
To account for the temporal coherence of the gravitational force relative to other forces driving the gas in the ISM, we compare the averaged gravitational acceleration of the gas to the corresponding average of the accelerations due the other forces, (the magnetic field and pressure gradients), using gas tracer particles.
This approach takes into account the fact that gravity is more likely than other forces to be coherent over space and throughout the lifetime of the cloud complex, being oriented along the gradient of the gravitational potential, as illustrated in Appendix~\ref{sec:time_evo}.
We compute the fraction of gravity-driven gas in the inflow towards high-density regions by selecting the gas for which gravity contributes more than half of the total integrated acceleration.

Overall, we find that the contribution of gravitational infall to the mass assembly of clouds is small, albeit non-zero. At large scales, the gravitational contribution is mostly due to the vertical stratification of the disk.
However, we also find that the contribution of gravity increases smoothly as the environment becomes denser and as time proceeds. The transition to a fully gravity-driven accretion flow on small scales would lie below our resolution limit. 

Our detailed findings are as follows:
\begin{enumerate}
    \item Gravity-driven gas can be found up to 100 pc away from the center of the gravitational potential of a large complex of clouds, but it is always subdominant on the scales investigated, implying that other physical processes dominate the structure and strength of the gas flows associated with the assembly of molecular cloud complexes in our simulation.
    
    \item About 10\% of the mass inflow onto the cold envelope of the cloud complex is gravity-driven. This is almost entirely due to gas infalling in the vertical direction and is thus a consequence of the large-scale gravitational potential, not the self-gravity of the gas.
    
    \item Gravity-driven gas contributes to 20\% of the mass assembly of cloud material, defined as regions where the density exceeds $2\times10^{-22}\  \gpcc$. At the moment of maximum compression (before photoionizing feedback begins to disrupt the cloud from inside), gravity-driven motions reach $\sim 45\%$ of the inflow at densities $\sim 10^{-21}\ \gpcc$. 
    
    \item The inflow within the clouds is increasingly dominated by the self-gravity of the gas as the density increases and as time proceeds, until internal feedback begins to disperse the densest gas.
    
    \item A transition to gravity-driven inflow may occur 
    beyond the largest densities than we probe, $\sim 2\times 10^{-21}\ \gpcc$.
 
    \item At the 10–100 pc scale, the contribution of gravity-driven motions to the linewidth associated with dense gas is negligible.

\end{enumerate}
~

Our results suggest that, in the flow regime we have simulated, with SN and photoionizing feedback and vertical large-scale gravity, but without stellar spiral arms, the assembly of molecular clouds is accomplished mostly by SN-driven motions, but the clouds manage to become nearly gravity-driven near the maximum density we resolve (at $\rho \sim 10^{-21}\ \gpcc$) at the time of maximum mass concentration. After that, photoionization-driven motions begin to disrupt the clouds from the inside.

Also, our results suggest a possible explanation to the observation that molecular clouds tend to be close to equipartition between the kinetic energy in nonthermal motions and their gravitational energy: that, rather from constituting a form of support, those motions consist of external compressions that contribute to the  assembly of the clouds. Therefore, the clouds become denser and more massive, thus increasing their gravitational energy, until it becomes dominant, leading to the onset of star formation which, through photoionizing radiation, erodes and disperses the clouds from the inside.

\section*{Acknowledgements}

The authors thank the two referees for their comments that helped clarify and improve the manuscript. 
They acknowledge Interstellar Institute's programs "II5: With Two Eyes", "II6" and "II7",
as well as the Paris-Saclay University's Institut Pascal for hosting discussions 
that triggered this work and nourished the development of the ideas behind it.
In particular, NB thanks Mordecai Mac-Low, Juan Soler, Blakesley Burkhart, Ioana Stelea, Marc-Antoine Mivilles-Deschênes, Javier Ballesteros-Paredes, Alessio Traficante, Guillaume Laibe, Benoît Commerçon and Henrik Beuther for fruitful discussions and interesting suggestions.
This research has received funding from the European Research Council
synergy grant ECOGAL (Grant: 855130). 
NB and JF acknowledge support from the ANR BRIDGES grant (ANR-23-CE31-0005). 
EVS acknowledges support from UNAM-PAPIIT grant IG100223.
Besides ECOGAL, RSK thanks for  financial support from the German Excellence Strategy via the Heidelberg Cluster of Excellence (EXC 2181 - 390900948) ``STRUCTURES'' and from the German Ministry for Economic Affairs and Climate Action in project ``MAINN'' (funding ID 50OO2206). 
Simulations were produced using the TGGC (Très Grand Centre de Calcul), through the GENCI allocation grant A0170411111.
We gratefully acknowledge support from the CBPsmn (PSMN, Pôle Scientifique de Modélisation Numérique) of the ENS de Lyon for the computing resources. The platform operates the SIDUS solution \citep{quemenerSIDUSSolutionExtreme2013} developed by Emmanuel Quemener.
NB and RSK acknowledge computing resources provided by the Ministry of Science, Research and the Arts (MWK) of the State of Baden-W\"{u}rttemberg through bwHPC and the German Science Foundation (DFG) through grants INST 35/1134-1 FUGG and 35/1597-1 FUGG, and also for data storage at SDS@hd funded through grants INST 35/1314-1 FUGG and INST 35/1503-1 FUGG.
The authors made use of the following tools: \textsc{Ramses}, Osyris, Numpy, Scipy, Numba, Matplotlib.

\section*{Data Availability}

The simulation data is available on the Galactica database (\url{http://www.galactica-simulations.eu/db/ISM/MASS_INFLOW/}) and the analysis scripts are available at the adress \url{https://git.brucy.fr/noe/mass_inflow_gravity}.


\bibliographystyle{mnras}
\bibliography{refs,global} 



~

\clearpage

\appendix

\section{Time evolution}
\label{sec:time_evo}

\begin{figure}[hb]
    \centering
    \includegraphics[width=\linewidth]{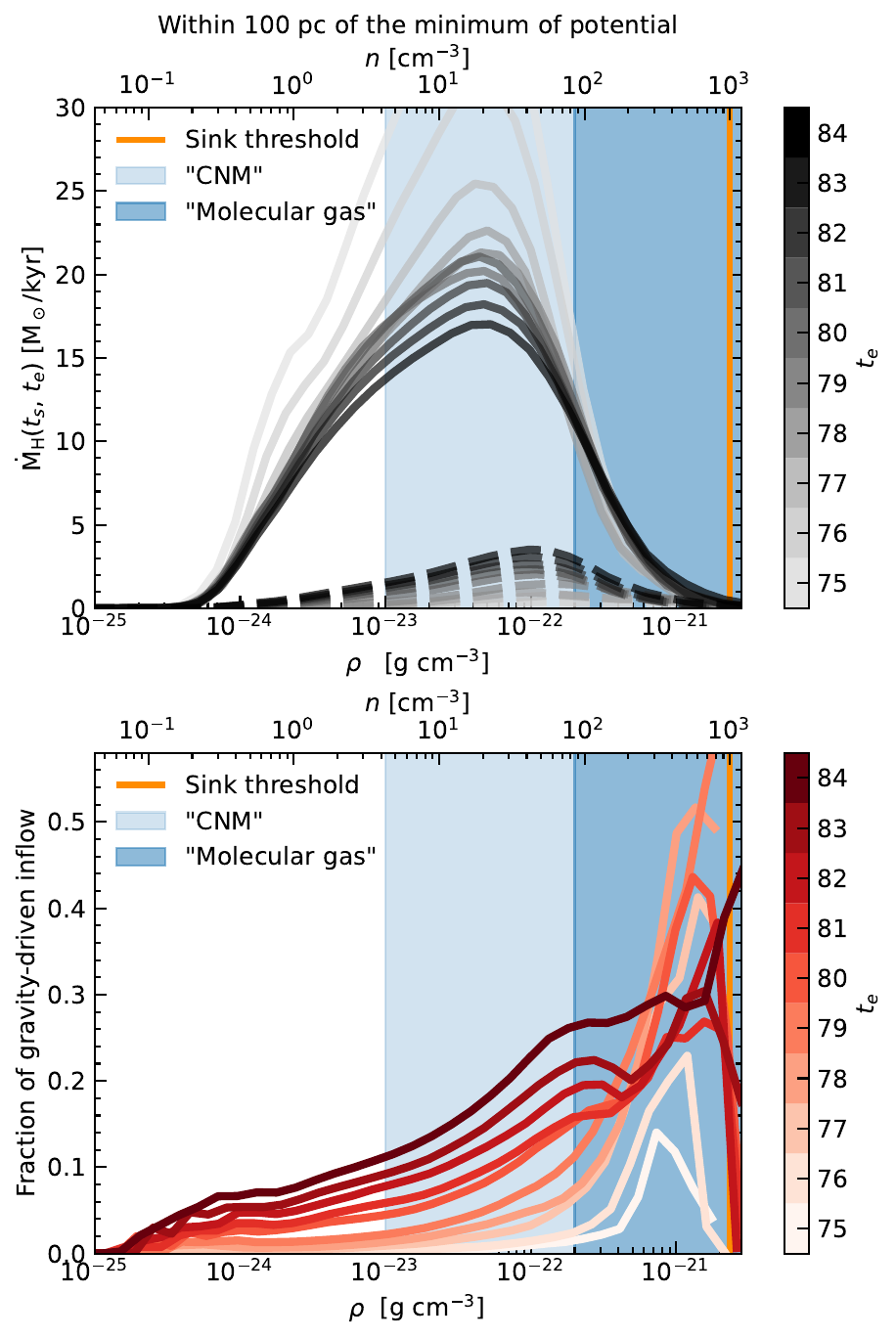}
    \caption{Top: Time evolution of the mass flow $\dot{\mathrm{M}}_H(t_s, t_e)$, but where the ending time $t_e$ varies from 75 to 84 Myr. The contribution of gravity-driven tracers is also plotted (dashed). Bottom: time evolution of the fraction of gravity-driven gas. }
    \label{fig:time_evo}
\end{figure}

An important feature of our study is that we integrate the accelerations over time to determine if the motion of tracer particle is gravity-driven or not.
Indeed, while turbulent force may dominate instant acceleration, its stochastic nature means that it may average out over time. This is not the case for gravitational forces, that are consistently pointing in the same direction over the lifetime of the large scale dense structures.
Figure~\ref{fig:time_evo} illustrates this effect, by looking at the mass inflow across density layers but at increasingly higher values of the integration time $t_e - t_s$, starting from 1 Myr to the fiducial value of 10 Myr. We do it by varying $t_e$ and keeping $t_s$ constant.
We see that the fraction of gravity-driven flow increases with the integration time. The increase is monotonous in the diffuse medium (densities lower than $2 \times10^{-22} \gpcc$ but reaches a peak at 45\% after 6 Myr of integration and then decreases in the denser parts. This is in part due to the formation of 2 sink particles inside the cloud complex and the occurrence of photoionizing feedback at t = 80 Myr.

\section{Cloud extraction}
\label{sec: clouds}

\begin{figure}[hb]
   \includegraphics[width=0.9\linewidth]{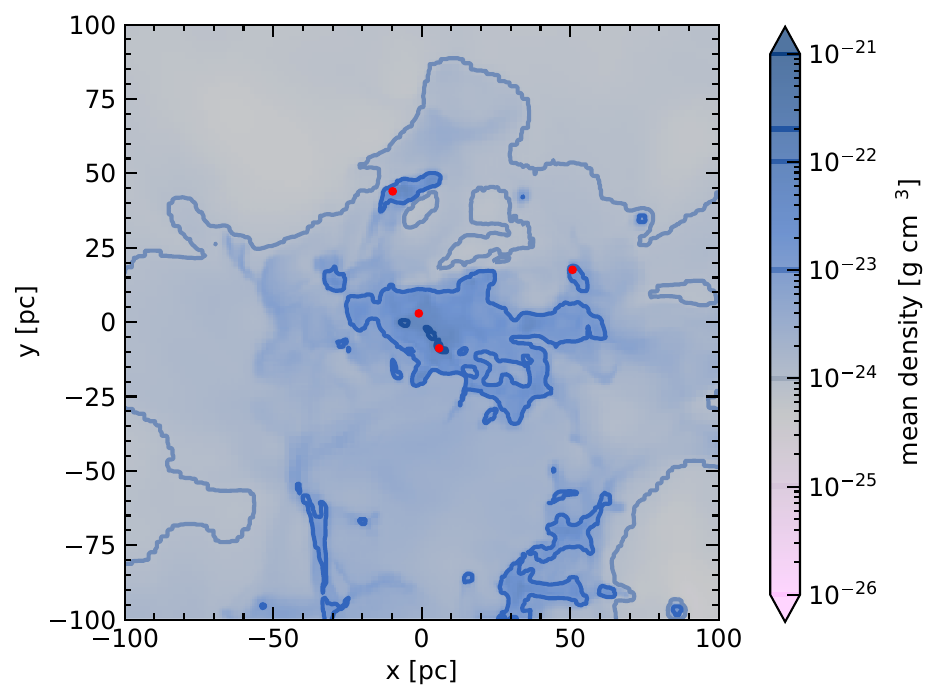}
    \includegraphics[width=0.9\linewidth]{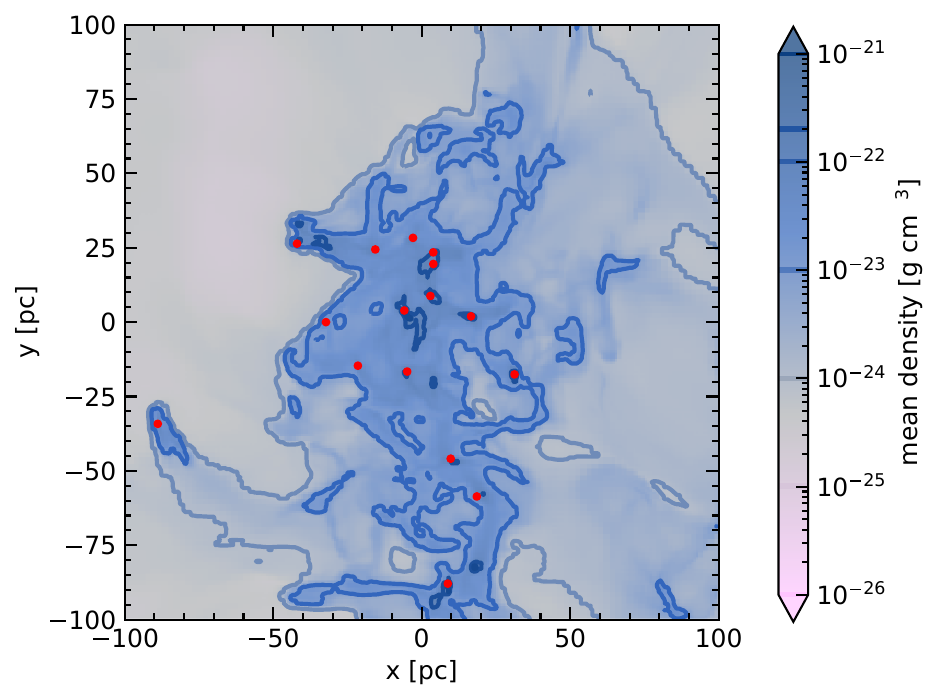}
    \caption{Position of the clouds in the midplane at $t_s$ (top) and $t_e$ (bottom). The red dots denotes the position of the peak density in the cloud. The background is the same as the left panel of Figure~\ref{fig:mass_inflow_density}.}
    \label{fig:clouds_pos}
\end{figure}

\begin{figure}
    \centering

    \includegraphics[width=0.8\linewidth]{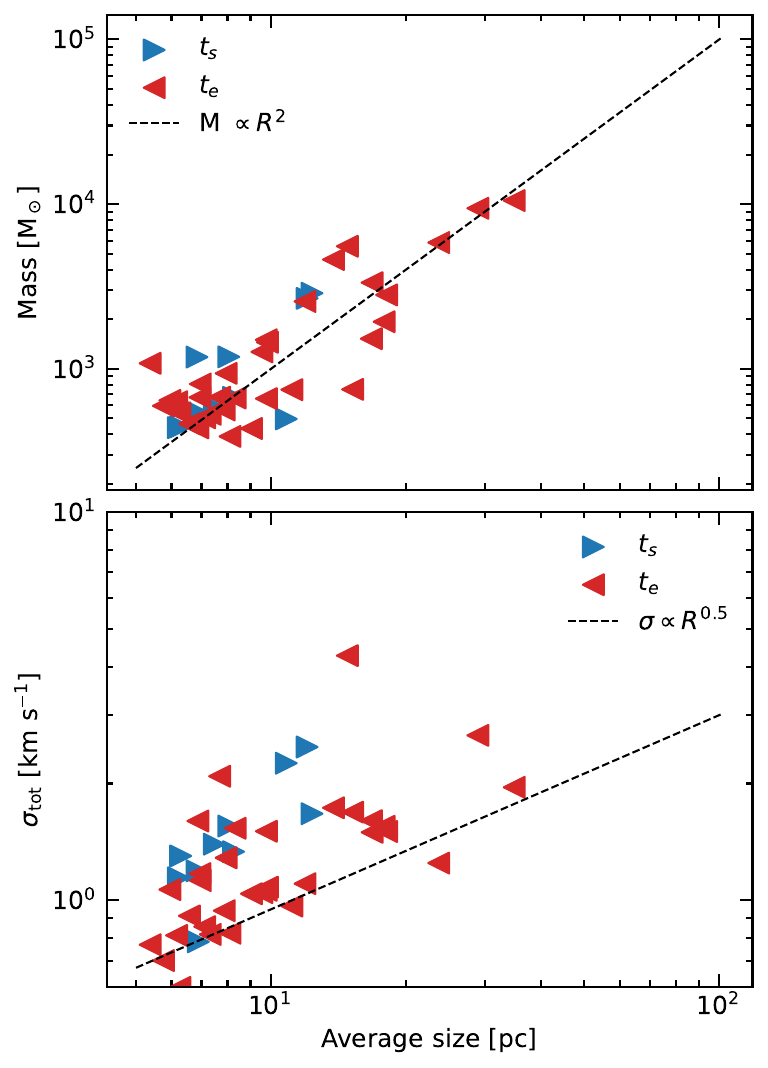}

 \caption{Size-mass (left) and size dispersion (right) relations of the clouds in the main cloud complex at $t_s$ and $t_e$. Dotted slope lines are not fits and are there to help reading the diagram.}
    \label{fig:clouds_rel}
\end{figure}

In order to extract the clouds of the simulation, we ran the ECOGAL wrapper \citep{colmanCloudPropertiesSpatial2024} around the HOP cloud detection algorithm \citep{eisensteinHOPNewGroupFinding1998}, with a threshold of $5\times 10^{-22}~\gpcc$, a peak factor of 2, a saddle factor of 1 and a minimum number of 100 cells. In Figure~\ref{fig:clouds_pos}, we indicate the location of the clouds within the cloud complex while in  Figure~\ref{fig:clouds_rel} we show the properties of the extracted clouds at the beginning and the end of the studied time-span. The extracted clouds have a wide variety of sizes, masses and velocity dispersions with values that are overall in line with observed and simulated clouds \citep{larsonTurbulenceStarFormation1981,miville-deschenesPhysicalPropertiesMolecular2017,colmanCloudPropertiesSpatial2024}.

\
\section{Tracer density drift}
\label{sec:tracer_drift}

\begin{figure}[h!]
    \centering
    \includegraphics[width=0.9\linewidth]{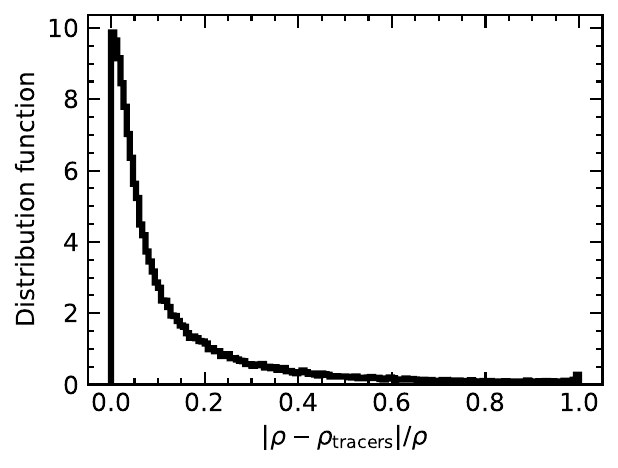}
    \caption{Mass-weighted distribution function of the relative error on the density of the mesh $\rho$ and the density $\rho_\mathrm{tracers}$ reconstructed through a cloud-in-cell algorithm based on the tracers' mass and position. Both were computed on a 256$^3$ sized-cube, corresponding to the minimum resolution in the simulation of $4$~pc at $t=t_e$. The error is computed in each cell of the cube, by comparing the  reconstructed density $\rho_\mathrm{tracers}$ with the actual density~$\rho$.}
    \label{fig:tracer_err}
\end{figure}

Velocity advected tracers are known to deviate from the gas flow over time \citep{priceComparisonGridParticle2010,genelFollowingFlowTracer2013}. In Figure~\ref{fig:tracer_err} we quantify this error by comparing the gas density with the density reconstructed through a cloud-in-cell interpolation from the tracers' mass and position, over a 256$^3$ sized-cube. That corresponds to the minimum resolution in the simulation of $4$~pc. 
In details, the cloud-in-cell algorithm considers a volume of (4 pc)$^3$ around each tracer. The mass of each tracer is then deposited onto the grid cells its surrounding volume intersect with, weighted by the volume of the intersection.
This allows to reconstruct a density from the tracers $\rho_\mathrm{tracers}$, which is then compared on a cell-by-cell basis to the actual density $\rho$.  In Figure~\ref{fig:tracer_err} we show the mass-weighted distribution of the relative error\[\mathrm{Err} = \dfrac{\vert\rho_\mathrm{tracers} - \rho \vert}{\rho}\] in each cell of this reconstructed cube.
Overall we find for 70\% of the tracers' mass, the error on the density is below 15\%, which indicates that the impact of the tracers' drift on our result is very limited.

\section{Effect of the accretion of the sink particles}
\label{sec:tracers_sink}

\begin{figure}

    \includegraphics[width=\linewidth]{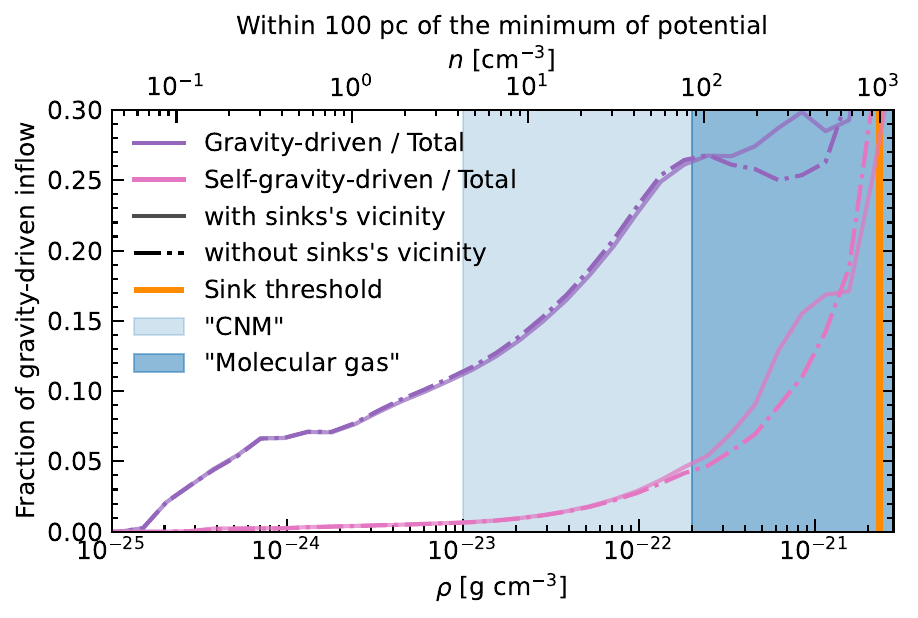}
      \includegraphics[width=\linewidth]{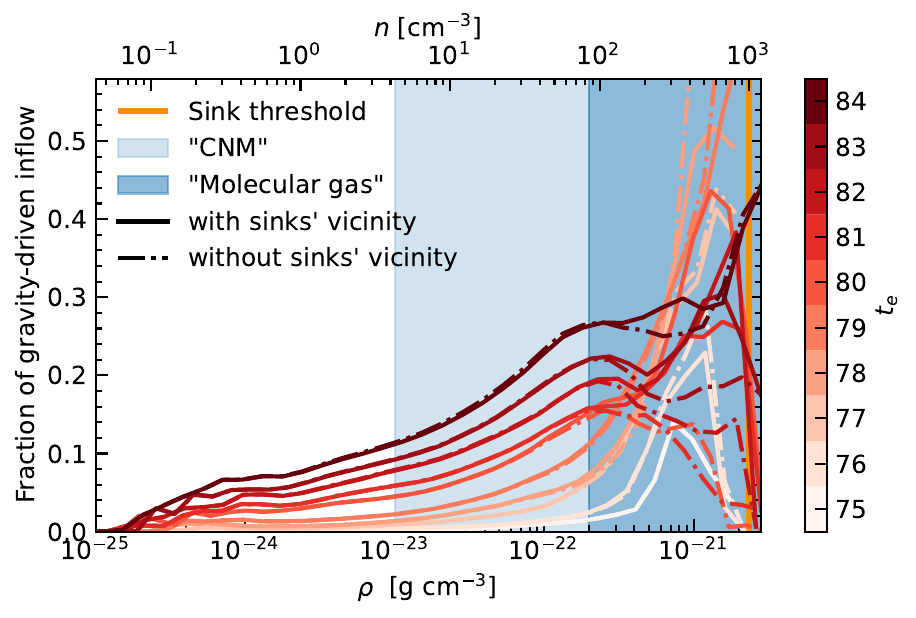}
    \caption{Top: Comparison of the mass fraction of gravity (purple) and self-gravity (pink) driven inflow across density layers when excluding the tracers within twice the accretion radius of a sink (dash-dotted lines) or not (plain lines).
    Bottom: Comparison  of gravity-driven inflow fraction for different ending time $t_e$ when excluding the tracers within twice the accretion radius of a sink (dash-dotted lines) or not (plain lines). }
    \label{fig:mass_inflow_nosink}
\end{figure}

In this section we quantify the impact of tracers near the sinks on our results. To this end, we recompute the fraction of the inflow which is gravity-driven and self-gravity-driven (shown in Figures~\ref{fig:mass_inflow_density}, right and \ref{fig:mass_inflow_sg}, respectively), but this time excluding the tracers closer than 8 pc (twice the accretion radius) of a sink. In Figure~\ref{fig:mass_inflow_nosink} (top), we see that including or not the tracers close to a sink has almost no influence on the result, except at intermediate number densities around $2 \times 10^2\ \mathrm{cm}^3$.
 Indeed, we note a slight decrease of the self-gravity-driven inflow fraction when removing the tracers around the sinks from the analysis. This may be a consequence of the accretion onto the sink, which causes the reduction of the thermal and ram pressure forces. It could also be because the gas close to a sink is affected by the gravity of the sink, which is also counted as a form of self-gravity.

 For completeness, we did the same exercise on all the studied snapshots, and compared with Figure~\ref{fig:time_evo}. Figure~\ref{fig:mass_inflow_nosink} (bottom) shows the fraction of gravity-driven inflow with and without the tracers closer than 8 pc to a sink particle for all snapshots. Overall, the effect is very small, except at high densities for times between 80 and 82 Myr, which is when most of the sinks form in the cloud complex.

\section{Gas density distribution}
\label{sec:pdf}

\begin{figure}
    \centering
    \includegraphics[width=0.98\linewidth]{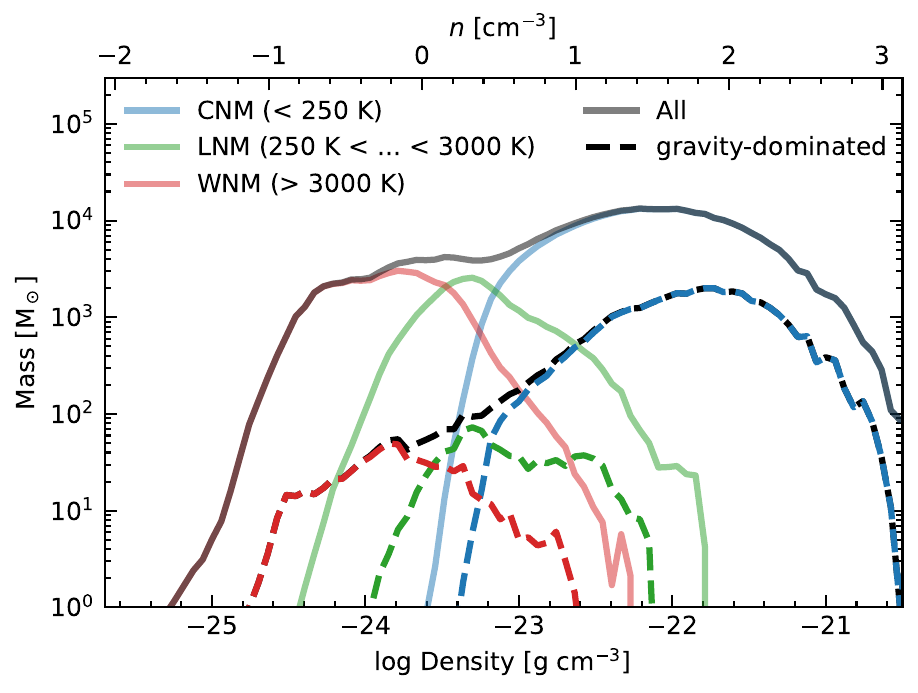}
    \caption{Mass-weighted density distribution function at $t=t_e$, with contribution from gravity-driven gas and the different phase of the gas.}
    \label{fig:pdf_200pc}
\end{figure}

Figure \ref{fig:pdf_200pc} shows the mass-weighted density distribution within 100 pc of the minimum of the potential. The contribution of the cold neutral medium (CNM), the lukewarm neutral medium (LNM) and the warm neutral medium (WNM), determined with the same temperature thresholds as in \cite{colmanRoleTurbulenceSetting2025}, are shown, as well as the contribution of the gravity-driven gas.
In details, the gas below 250 K is considered to be CNM, while the gas above 3000 K is considered WMN. The gas in between is LMN. The mass of gravity-dominated gas in each cell is computed from the fraction of gravity-dominated tracers in that cell.
There is not a clear power-law tail in the cold gas, which consistent with our conclusion that gravity does not drive the mass flows at the scales probed by our simulation.

\newpage

\end{document}